\documentstyle[12pt] {article}
\newcommand{\beq}{\begin{equation}}
\newcommand{\eeq}{\end{equation}}
\newcommand{\beqa}{\begin{eqnarray}}
\newcommand{\eeqa}{\end{eqnarray}}
\newcommand{\ba}{\begin{array}}
\newcommand{\ea}{\end{array}}

\begin{document}

\begin{center}
{\bf \large Order and Chaos in Roto--Vibrational \\
States of Atomic Nuclei}
\footnote{This work has been partially supported by the Ministero \\
dell'Universit\`a e della Ricerca Scientifica e Tecnologica (MURST).}

\end{center}

\vspace{0.5 cm}

\begin{center}
{\bf V.R. Manfredi}
\footnote{Author to whom all correspondence and
reprint requests should be addressed. \\
E--Mail: VAXFPD::MANFREDI, MANFREDI@PADOVA.INFN.IT} \\
Dipartimento di Fisica ``G. Galilei" dell'Universit\`a
di Padova, \\
INFN, Sezione di Padova, \\
Via Marzolo 8, I 35131 Padova, Italy \\
\vskip 0.3 truecm
Interdisciplinary Laboratory, SISSA,\\
Strada Costiera 11, 34014 Trieste, Italy\\
\end{center}

\vskip 0.5 truecm

\begin{center}
{\bf L. Salasnich}  \\
Dipartimento di Fisica "G. Galilei" dell'Universit\`a
di Padova, \\
INFN, Sezione di Padova, \\
Via Marzolo 8, I 35131 Padova, Italy \\
\vskip 0.3 truecm
Departamento de Fisica Atomica, Molecular y Nuclear\\
Universidad "Complutense" de Madrid, \\
Av. Complutense, E 28040 Madrid, Espana
\end{center}

\vskip 0.5 truecm

\newpage

\begin{center}
{\bf Abstract}
\end{center}
\vskip 0.5 truecm
\par
Using a classical analytical criterion (that of curvature) and
numerical results (Poincar\`e sections and spectral statistics),
a transition order--chaos--order in the roto--vibrational
model of atomic nuclei has been shown. Numerical
calculations were performed for some deformed nuclei.

\vskip 2. truecm

\begin{center}
\end{center}

\newpage

\par
{\bf 1. Introduction}
\vskip 0.5 truecm
\par
In the last few years many authors have shown great interest in the
so--called "quantum chaos": the properties of quantal systems which are
chaotic in the semiclassical limit $\hbar \to 0$ [1,2,3].
\par
In atomic nuclei the coexistence of regular, chaotic and collective
states makes the problem quite intricate. In order to disentangle it,
many models have been used (see, for example [4,5,6] and references
quoted therein). The aim of this paper is to study
the transition from ordered to chaotic states in the roto--vibrational
model of atomic nuclei. This model was introduced by Bohr and Mottelson [7]
and discussed in great detail in [8,9,10].

\vskip 0.5 truecm
\par
{\bf 2. The Model}
\vskip 0.5 truecm
\par
Since the model has been amply described in [8], in this section we
limit ourselves to reporting only a few basic formulae.
\par
The hamiltonian is:
\beq
H=E_{vib}+E_{rot},
\eeq
where:
\beq
E_{vib}={1\over 2}B({\dot a}_0^2+2{\dot a}_2^2)+V(a_0,a_2),
\eeq
\beq
E_{rot}={1\over 2}\sum_{k=1}^3 \omega_k^2 J_k(a_0,a_2),
\eeq
with:
\beq
V(a_0,a_2)=\sum_{m,n}c_{mn}(a_0^2+2a_2^2)^m a_0^n(6a_2^2-a_0^2)^n.
\eeq
The parameters $a_0$ and $a_2$ are connected to the deformation
$\beta$ and asymmetry $\gamma$ by the standard relations [8]:
\beq
a_0=\beta \cos{\gamma}, \;\;\; a_2={\beta\over \sqrt{2}}\sin{\gamma}.
\eeq
In terms of the new variables the components of the moment of inertia
are [7]:
\beq
J_k=4B\beta^2 \sin^2{(\gamma -{2\pi\over 3}k)}.
\eeq
If the nucleus has an axially symmetric deformation:
\beq
\omega_1=\omega_2={\omega\over \sqrt{2}}, \;\;\; \omega_3=0,
\eeq
and if we take only the first terms of (4), the hamiltonian (1) can be
written:
\beq
H={1\over 2}B({\dot a}_0^2+2{\dot a}_2^2)+V(a_0,a_2)
+{1\over 2}B\omega^2 (3a_0^2+2a_2^2),
\eeq
where:
\beq
V(a_0,a_2)={1\over 2}C_2(a_0^2+2a_2^2)+
\sqrt{2\over 35}C_3a_0(6a_2^2-a_0^2)+{1\over 5}C_4(a_0^2+2a_2^2)^2+V_0,
\eeq
with $V_0$ a parameter chosen to have the minimum of the potential equal
to zero. As discussed in [8],
the presence of bound states in atomic nuclei leads
to a value of $C_4>0$, whereas for $C_3$ a positive value corresponds to
a prolate shape, a negative value to an oblate shape. Similarly $C_2$ may
also be either positive or negative.
\par
The shape of the nuclear potential $V(a_0,a_2)$ is a function of $C_2$
and $\chi =C_3^2/(C_2 C_4)$. For $C_2>0$, and $0<\chi<56/9$
the nucleus is spherical; for $56/9<\chi<7$ the nucleus is spherical in
the ground state (g.s.) and deformed in the excited states (e.s.);
for $\chi >7$ it is deformed in the g.s. and spherical in the e.s;
for $C_2<0$ it is deformed in the g.s. and $\gamma$--unstable
in the e.s.
\par
Near the equilibrium ($\bar{a_0}=\beta_0$, $\bar{a_2}=0$),
$V(a_0,a_2)$ can be written:
\beq
\tilde{V}(a_0,a_2)={\tilde{C_0}\over 2}(a_0-\beta_0)^2+{\tilde
C_2}a_2^2.
\eeq
\par
In order to calculate the equilibrium configuration the following system
must be solved:
\beq
{\partial V\over \partial a_0}|_{\beta_0,0}=\beta_0
(C_2-3\sqrt{2\over 35}C_3\beta_0 +{4\over 5}C_4\beta_0^2)=0,
\eeq
\beq
{\partial^2 V\over \partial a_0^2}|_{\beta_0,0}=
C_2-6\sqrt{2\over 35}C_3\beta_0 +{12\over 5}C_4\beta_0^2 =
{\partial^2 \tilde{V}\over \partial a_0^2}|_{\beta_0,0}=\tilde{C_0},
\eeq
\beq
{1\over 2}{\partial^2 V\over \partial a_2^2}|_{\beta_0,0}=
C_2+6\sqrt{2\over 35}C_3\beta_0 +{12\over 5}C_4\beta_0^2 =
{\partial^2 \tilde{V}\over \partial a_2^2}|_{\beta_0,0}=\tilde{C_2}.
\eeq
\par
In roto--vibrational nuclei the parameters $\tilde{C_0}$ and
$\tilde{C_2}$ are simply connected to some experimental quantities:
\beq
E_{\beta}=\hbar \omega_{\beta}=\hbar\sqrt{\tilde{C_0}\over B},
\;\;\;\; E_{\gamma}=\hbar \omega_{\gamma}=
\hbar \sqrt{\tilde{C_2}\over B},
\eeq
and:
\beq
\tilde{C_0}={E_{\beta}^2\over 3\epsilon \beta_0^2}, \;\;\;\;
\tilde{C_2}={E_{\gamma}^2\over 3\epsilon \beta_0^2},
\eeq
where $\epsilon=\hbar^2/J_0$ is taken from the experimental energy of
the first rotational state and $\beta_0$ is the equilibrium deformation
of the nucleus, which may be obtained from the $B(E_2,2^+\to 0^+)$ value
[8]. Finally:
$$
C_2=-{1\over 18\epsilon \beta_0}(3E_{\beta}^2-E_{\gamma}^2),
$$
\beq
C_3=\sqrt{32\over 5}{E_{\gamma}^2\over 27 \epsilon \beta_0^2},
\eeq
$$
C_4={5\over 72 \epsilon \beta_0^3}(3E_{\beta}^2-E_{\gamma}^2).
$$
\par
As a simple application of the above procedure, the nuclear potentials
have been calculated for $^{164}Dy$ and $^{166}Er$ (see
Figures 1,2). The numerical values of the parameters $\beta_0$,
$\epsilon$, $E_{\beta}$, $E_{\gamma}$, taken from the experimental data,
are shown in Table 1.

\vskip 0.5 truecm
\par
{\bf 3. The onset of chaos by curvature criterion}
\vskip 0.5 truecm
\par
As is well known, the transition order--chaos in systems with two
degrees of freedom may be studied by the curvature criterion [11,12].
It is however important to point out that {\it in general} the curvature
criterion guarantees only a {\it local instability} and should therefore
be combined with the Poincar\`e sections [14]. For a fuller discussion of
this point see [13]. As mentioned in
the previous section, the total hamiltonian can be written:
\beq
H={1\over 2}B({\dot a}_0^2+2{\dot a}_2^2)+W(a_0,a_2),
\eeq
where:
\beq
W(a_0,a_2)=V(a_0,a_2)+{1\over 2}B\omega^2 (3a_0^2+2a_2^2)
\eeq
is the effective potential. As a function of $C_2$, $C_3$, $C_4$, (18)
becomes:
\beq
W(a_0,a_2)={1\over 2}C_2(a_0^2+2a_2^2)+
\sqrt{2\over 35}C_3a_0(6a_2^2-a_0^2)+{1\over 5}C_4(a_0^2+2a_2^2)^2+V_0
+{1\over 2}B\omega^2 (3a_0^2+2a_2^2).
\eeq
Owing to the symmetry properties of the effective potential $W$,
our study may be restricted to the case $W(a_0,a_2=0)$ [8].
To apply the above criterion to our system,
the sign of the curvature $K$ can be studied by solving the equation:
\beq
K(a_0)={\partial^2 W\over \partial a_0^2}(a_0,a_2=0)
={12\over 5}C_4 a_0^2-6\sqrt{2\over 35}C_3a_0+(C_2+3B\omega^2)=0,
\eeq
whose discriminant $\Delta$ is given by:
\beq
\Delta ={72\over 35}C_2 C_4 (\chi -{14\over 3}-14{B\omega^2\over C_2}).
\eeq
If $\Delta \leq 0$ the curvature $K$ is always positive and the motion
is regular; if $\Delta >0$ there is a region of negative curvature and
the motion may be chaotic.
\par
For $C_2>0$, $0 <\chi < 14/3$ (spherical nuclei)
the curvature is positive and therefore the motion is regular for all
$\omega$. For $\chi \geq 14/3$ (spherical and deformed nuclei) and:
\beq
0\leq \omega < \sqrt{ {C_2\over 14 B}(\chi -{14\over 3})}
\eeq
the curvature is negative and chaotic motion may appear.
\par
For $C_2<0$ ($\gamma$--unstable nuclei) there is a region with negative
curvature for $0\leq \omega < \omega_c$, where:
\beq
\omega_c=\sqrt{ {C_2\over 14 B}(\chi -{14\over 3})}
\eeq
is the critical frequency of the system. As a function of $\epsilon$ and
$\beta_0$ the critical energy can be written:
\beq
\hbar \omega_c=\sqrt{(\chi -{14\over 3})
{3\epsilon \beta_0^2 C_2\over 14} }
\eeq
and the critical angular momentum $I_c$ is:
\beq
I_{c}=\hbar {1\over \epsilon}
\sqrt{(\chi -{14\over 3}) {3\epsilon \beta_0^2 C_2\over 14} }.
\eeq
It is perhaps noteworthy that the shape of the effective potential
$W(a_0,0)$ changes drastically as a function of $\omega$ (see Fig. 3).
If $\omega$ increases there is a transition from chaos to order:
the region of chaotic motion decreases and becomes zero for
$\omega >\omega_c$.
For a fixed value of the parameter $\chi$, the region of chaotic motion is
limited, on the line $K(a_0)=0$, by the two branches defined by:
\beq
a_0^{\pm}={5\over 4C_4}\big[ \sqrt{2\over 35}C_3\pm
\sqrt{ {2\over 35}C_2C_4(\chi -{14\over 3}-{14B\omega^2\over C_2})
}\big].
\eeq
Table 3 shows, for some nuclei,
the energies $E^-_{chaos}$ and $E^+_{chaos}$,
which limit the chaotic region for $\omega =0$, and also the
critical angular momentum $I_c$.

\vskip 0.5 truecm
\par
{\bf 4. Numerical study of the order--chaos transition}
\vskip 0.5 truecm
\par
As mentioned in the previous section, the curvature criterion is able to
characterize the local behaviour of the system (for example the local
instability) and may give only a rough signature of the global
properties (e.g. the global instability) [13]. As is well known, a very
useful tool for the study of global properties is provided by the
Poincar\`e sections [14]. With this aim the classical trajectories have
been calculated by a fourth order Runge--Kutta method. In order to avoid
numerical errors connected to the use of finite temporal intervals, a
first--order interpolation has been used [15].
\par
The Hamilton equations of the systems are:
$$
{\dot a_0}=B p_{0},
$$
$$
{\dot a_2}=2B p_{2},
$$
\beq
{\dot p_{0}}=-C_2a_0-2\sqrt{2\over 35}C_3(3a_2^2-3a_0^2)-
{4\over 5}C_4a_0(a_0^2+2a_2^2)-3B\omega^2a_0,
\eeq
$$
{\dot p_{2}}=-2C_2a_2-12\sqrt{2\over 35}C_3a_0a_2-
{8\over 5}C_4a_2(a_0^2+2a_2^2)-2B\omega^2a_2,
$$
where $p_0$ and $p_2$ are the conjugate momenta:
\beq
p_{0}=B\dot{a_0}, \;\;\;\; p_{2}=2B\dot{a_2}.
\eeq
Figure 4 shows the Poincar\`e sections for $^{160}Gd$ at the energy
$5.5$ MeV and for different values of rotational frequency. The figure
clearly shows a chaos--order transition as the frequency $\omega$
increases. In Figure 5 for $^{166}Er$ the Poincar\`e sections are shown
for different values of the energy and rotational frequency $\omega =0$.
As can be seen, there is a chaos--order transition, albeit not so sharp
as in the previous case.
\par
It is well known that the fluctuation properties of
quantal systems with underlying classical chaotic behaviour and
time--reversal symmetry are in agreement with the predictions of
the Gaussian Orthogonal Ensemble (GOE), and that quantum analogs of
classically integrable systems
display the characteristics of Poisson statistics [4,6].
\par
For deformed nuclei, like those of the rare--earth region, it is not
easy to obtain the correct energy levels, because the potential energy
is an asymmetric triple well. To avoid the problems related
to the tunneling effects, a numerical method based on the
formulation of quantum mechanics using the euclidean
path integral [16] should be used. The energy spectrum
is then mapped into one with quasi--uniform level density
by means of the local unfolding procedure
described in detail in reference [17].
\par
In Figure 6 the spectral statistics
$P(s)$ and $\Delta_{3}$ [18,19] are plotted for $^{166}Er$. These
statistics confirm the classical results: for energies above
the saddle energy, about $4$ MeV, there is
prevalently chaotic behaviour; for higher energies, about $15$ MeV,
there is mixed behaviour with a predominance of regular classical
trajectories. The non--universal behaviour of $\Delta_3(L)$
for large values of $L$, not predicted by GOE, has been explained by
Berry [20] using the semiclassical quantization.

\vskip 0.5 truecm
\par
{\bf 5. Conclusions}
\vskip 0.5 truecm
\par
We have shown by combining analytical results (the curvature criterion)
with numerical ones (the Poincar\`e sections and spectral statistics)
that, in the roto--vibrational model of atomic nuclei, an
order--chaos--order transition occurs as a function of the energy.
Our results are in good agreement with those of [12], but,
unlike the authors of [12], we have also calculated the energy ranges of
the chaotic regions (see Table 3) and the chaos--order transition
which occurs as a function of the rotational frequency $\omega$.

\vskip 0.5 truecm
\par
The authors are greatly indebted to M. Rosa--Clot and S. Taddei
for the provided numerical data of energy levels.

\newpage

{\bf REFERENCES}
\vskip 0.5 truecm

[1] M.C. Gutzwiller: {\it Chaos in Classical and Quantum Mechanics}
(Springer--Verlag, Berlin, 1990)

[2] M. Berry: in {\it Quantum Chaos}, Ed. H.A. Cerdeira, R. Ramaswamy,
M.C. Gutzwiller, G. Casati (World Scientific, New York, 1991)

[3] {\it From Classical to Quantum Chaos}, Ed. G.F. Dell'Antonio, S.
Fantoni, V.R. Manfredi, Conference Proceedings SIF, vol. 41
(Editrice Compositori, Bologna, 1993)

[4] O. Bohigas, H.A. Weidenm\"uller: {\it Ann. Rev. Nucl. Part. Sci.}
{\bf 38}, 421 (1988)

[5] B.R. Mottelson: {\it Nucl. Phys.} A {\bf 557}, 717c (1993)

[6] M.T. Lopez--Arias, V.R. Manfredi, L. Salasnich:
{\it La Rivista del Nuovo Cimento} {\bf 17}, n. 5 (1994)

[7] A. Bohr, B. Mottelson: {\it Nuclear Structure}, vol. 2 (Benjamin,
London, 1975)

[8] J.M. Eisenberg, W. Greiner: {\it Nuclear Models}, vol. 1 (North Holland,
Amsterdam, 1970)

[9] A. Faessler, W. Greiner: {\it Zeit. Phys.} {\bf 168}, 425 (1962);
A. Faessler, W. Greiner R.K. Sheline: {\it Nucl. Phys.} {\bf 70}, 33 (1965)

[10] U. Mosel, W. Greiner: {\it Zeit. Phys.} {\bf 217}, 256 (1968)

[11] M. Toda: {\it Phys. Lett.} A {\bf 48}, 335 (1974)

[12] Yu. Bolotin, V.Yu. Gonchar, E.V. Inopin, V.V. Levenko, V.N. Tarasov,
N.A. Chekanov: {\it Sov. I. Part. Nucl.} {\bf 20}, 372 (1989)

[13] G. Benettin, R. Brambilla, L. Galgani: {\it Physica} A {\bf 87},
381 (1977)

[14] H. Poincar\`e: {\it New Methods of Celestial Mechanics}, vol. 3, ch.
27 (Transl. NASA Washington DC 1967)

[15] M. Henon: {\it Physica} D {\bf 5}, 412 (1982)

[16] M. Rosa-Clot, S. Taddei: {\it Phys. Rev.} C {\bf 50}, 627 (1994);
M. Rosa-Clot, S. Taddei: {\it Phys. Lett.} A {\bf 197}, 1 (1995)

[17] V.R. Manfredi: {\it Lett. Nuovo Cim.} {\bf 40}, 135 (1984)

[18] F.J. Dyson, M.L. Metha: {\it J. Math, Phys.} {\bf 4}, 701 (1963)

[19] O. Bohigas, M.J. Giannoni: Ann. Phys. (N.Y.) {\bf 89}, 393 (1975)

[20] M. Berry: {\it Proc. Roy. Soc. Lond.} A {\bf 400}, 229 (1985)

\newpage

{\bf TABLE CAPTIONS}
\vskip 0.5 truecm

Table 1: Numerical values of the parameters $\beta_0$, $\epsilon$,
$E_{\beta}$, $E_{\gamma}$, taken from the experimental data for some
even--even nuclei. The parameters are defined in the text.

Table 2: Numerical values of the parameters $C_2$, $C_3$, $C_4$, $V_0$
defining the effective potential $W$.

Table 3: Numerical values of the energies which delimit the chaotic
region for $\omega =0$ and the critical angular momentum.

\newpage

{\bf FIGURE CAPTIONS}
\vskip 0.5 truecm

Figure 1: Nuclear potential for $^{164}Dy$.

Figure 2: Nuclear potential for $^{166}Er$.

Figure 3: Effective nuclear potential for $^{164}Dy$ as a function of the
rotational frequency $\omega$; from left to right:
$\hbar \omega=0$ MeV, $\hbar \omega=0.5$ MeV, $\hbar \omega =1$ MeV.

Figure 4: The Poincar\`e sections for $^{160}Gd$ at the energy
$5.5$ MeV and for different values of rotational frequency;
from the top: $\hbar \omega=0$ MeV, $\hbar \omega=0.5$ MeV,
$\hbar \omega =1$ MeV.

Figure 5: The Poincar\`e sections for $^{166}Er$ at the
rotational frequency $\omega =0$ and for different values of the energy:
(a) $E=1$ MeV, (b) $E=6$ MeV, (c) $E=9$ MeV, (d) $E=12$ MeV.

Figure 6: Spectral statistics $P(s)$ and $\Delta_3(L)$ for $^{166}Er$
at the rotational frequency $\omega =0$ for different energy regions:
$2\leq E\leq 6$ MeV (below) and for $13\leq E\leq 17$ MeV (above).
The solid line is the GOE statistic curve and the dashed line is the
Poisson one.

\newpage

\begin{center}
\begin{tabular}{|ccccc|} \hline\hline
$Nucleus$ & $\beta_0$ & $\epsilon$ (MeV) & $E_{\beta}$ (MeV) &
$E_{\gamma}$ (MeV)\\ \hline
$^{160}Gd$ & 0.47 & 0.022 & 1.80 & 0.96\\
$^{164}Dy$ & 0.41 & 0.021 & 1.76 & 0.73\\
$^{166}Er$ & 0.33 & 0.023 & 1.14 & 0.75\\
$^{230}Th$ & 0.23 & 0.015 & 0.63 & 0.76\\
$^{238}Ur$ & 0.28 & 0.014 & 0.99 & 1.05\\
$^{240}Pu$ & 0.28 & 0.013 & 0.86 & 0.92\\ \hline\hline
\end{tabular}
\end{center}
\vskip 0.6 truecm
{\bf Table 1}

\newpage

\begin{center}
\begin{tabular}{|ccccc|} \hline\hline
Nucleus & $C_2$ (MeV) & $C_3$ (MeV) & $C_4$ (MeV) &
$V_0$ (MeV)\\ \hline
$^{160}Gd$ & -100.64 &  37.57 &  668.01 & 5.56\\
$^{164}Dy$ & -138.41 &  34.97 & 1155.67 & 5.69\\
$^{166}Er$ &  -73.77 &  64.96 & 1138.52 & 1.89\\
$^{230}Th$ &  -43.56 & 229.67 & 2260.62 & 0.44\\
$^{238}Ur$ &  -94.54 & 334.20 & 3276.70 & 1.56\\
$^{240}Pu$ &  -73.64 & 281.53 & 2664.58 & 1.20\\ \hline\hline
\end{tabular}
\end{center}
\vskip 0.6 truecm
{\bf Table 2}
\vskip 0.5 truecm

\newpage

\begin{center}
\begin{tabular}{|cccc|} \hline\hline
$Nucleus$ & $E_{chaos}^-$ (MeV) & $E_{chaos}^+$ (MeV) & $I_c (\hbar)$ \\
\hline
$^{160}Gd$ & 2.36 & 5.36 & 31.86 \\
$^{164}Dy$ & 2.51 & 5.59 & 33.31 \\
$^{166}Er$ & 0.83 & 1.89 & 18.79 \\
$^{230}Th$ & 0.15 & 0.44 & 13.28 \\
$^{238}Ur$ & 0.61 & 1.55 & 23.88 \\
$^{240}Pu$ & 0.46 & 1.19 & 21.97 \\ \hline\hline
\end{tabular}
\end{center}
\vskip 0.6 truecm
{\bf Table 3}

\end{document}